# Demonstration of negative refraction induced by synthetic gauge fields

*Yihao Yang[1,2,3,#], Yong Ge[4,#], Rujiang Li[2,3,#], Xiao Lin[1], Ding Jia[4], Yi-jun Guan[4], Shou-qi Yuan[4], Hong-xiang Sun[4,\*], Yidong Chong[2,3,\*], and Baile Zhang[2,3,\*]*

[1]Interdisciplinary Center for Quantum Information, State Key Laboratory of Modern Optical Instrumentation, ZJU-Hangzhou Global Science and Technology Innovation Center, Zhejiang University, Hangzhou 310027, China.

[2]Division of Physics and Applied Physics, School of Physical and Mathematical Sciences, Nanyang Technological University, 21 Nanyang Link, Singapore 637371, Singapore.

[3]Centre for Disruptive Photonic Technologies, The Photonics Institute, Nanyang Technological University, 50 Nanyang Avenue, Singapore 639798, Singapore.

[4]Research Center of Fluid Machinery Engineering and Technology, School of Physics and Electronic Engineering, Jiangsu University, Zhenjiang 212013, China.

#These authors contributed equally to this work.

*Correspondence to: (H. S.) jsdxshx@ujs.edu.cn; (Y. C.) yidong@ntu.edu.sg; (B. Z.) blzhang@ntu.edu.sg.

**Abstract**

The phenomenon of negative refraction generally requires negative refractive indices or phase discontinuities, which can be realized using metamaterials or metasurfaces. Recent theories have proposed a novel mechanism for negative refraction based on synthetic gauge fields, which affect classical waves as if they were charged particles in electromagnetic fields, but this has not hitherto been demonstrated in experiment. Here, we report on the experimental demonstration of gauge-field-induced negative refraction in a twisted bilayer acoustic metamaterial. The bilayer twisting produces a synthetic gauge field for sound waves propagating within a projected two-dimensional geometry, with the magnitude of the gauge field parameterized by the choice of wavenumber along the third dimension. Waveguiding with backward propagating modes is also demonstrated in a trilayer configuration that implements strong gauge fields. These results provide an alternative route to achieving negative refraction in synthetic materials.

Light, sound, and other waves can change direction when passing between two media with different refractive indices—the well-known phenomenon of refraction. Typically, the incident and refracted waves lie on opposite sides of the interface normal, a situation referred to as "positive" refraction. Decades ago, Veselago raised the possibility of "negative" refraction, whereby the refracted wave lies on the same side of the interface normal as the incident wave[1]. Following extensive theoretical debates[2-5], the concept of negative refraction is now widely accepted, and the phenomenon has been convincingly demonstrated in bulk metamaterials that exhibit negative effective refractive indices[6-10]. Moreover, it has recently been found that metasurfaces with properly-engineered phase discontinuities can exhibit negative refraction[11,12]. Aside from its significance for fundamental science, negative refraction has promising applications for optical lenses[13,14] and invisibility cloaks[15].

Several recent studies have explored using synthetic gauge fields as a novel way to manipulate classical waves, different from earlier approaches involving bulk metamaterials and metasurfaces, Gauge fields arise in numerous settings in modern physics; for instance, in the quantum mechanical description of a charged particle, the vector potential $A$ appears as a gauge field added to the canonical momentum. Although classical waves like light or sound carry no charge and thus do not couple to a real gauge field, it is possible to engineer structures or modulations that produce the effects of synthetic gauge fields[16-29]. Previous studies have shown how synthetic gauge fields can generate effective magnetic fields and associated phenomena[16,18,27,30-43]. For example, a photonic or acoustic crystal with strain engineering can produce a spatially-dependent gauge field $A$, whose resulting effective magnetic field $B=\nabla\times A$ gives rise to Landau levels and quantum-Hall-like edge states[19,21-23,27]. Likewise, a photonic lattice can be dynamically modulated to produce an effective magnetic field giving rise to topologically protected edge states[16].

However, the gauge field itself can be a powerful tool for wave manipulation even if the associated magnetic field is zero (i.e., $\nabla\times A=0$)[20,26,29,44,45]. For instance, a spatially uniform gauge field can shift the dispersion relation in momentum space[20,26,29], a phenomenon similar to the relativistic Fresnel drag effect, in which light-dragging by a moving medium leads to a displacement of the isofrequency dispersion contour[46]. According to a recent prediction, when a wave is incident from a conventional medium to a medium with a uniform gauge field of sufficient strength, negative refraction can occur[29]. This differs from previous ways of achieving negative refraction since it does not rely on bulk negative refractive indices in one of the media, or phase discontinuities along the interface. However, there has been no experimental demonstration of this phenomenon thus far.

Here, we report on the experimental demonstration of negative refraction induced by a gauge field acting as a non-reciprocal "one-way mirror". We also show how gauge fields can be used to form waveguides that support backward-propagating modes whose phase velocity and group velocity point in opposite directions. These experiments are carried out in twisted bilayer and trilayer acoustic metamaterials, whose gauge fields in the *x-y* plane are effectively parameterized by the choice of wavenumber along the third spatial axis.

The design of the twisted bilayer acoustic metamaterial is shown in Fig. 1(a). The bottom and top layers are three-dimensional (3D) structures consisting of alternating hard solid plates with thickness $w$=3 mm separated by air gaps with thickness $w_b$=4.1 mm. Because of the deeply subwavelength features, the two layers can be treated as a continuous effective medium. The top layer is twisted around the *x*-axis by an angle of $\varphi$. By selecting a fixed $k_z$, the 3D geometry can be reduced to a two-dimensional (2D) geometry in the *xy* plane. The isofrequency contour for the bottom layer is a 2D circle of radius $k_0$ (where $k_0$ is the wavevector in air) centred at the origin of momentum space, as shown in Fig. 1(b). The corresponding effective refractive index is $n_1$=1. The isofrequency contour for the top layer is an ellipse with major axis $k_0/\sin(\varphi)$ and minor axis $k_0$, meaning that the *x* (*y*) component of the corresponding effective refractive index is $n_x$=1 [$n_y$=1/sin($\varphi$)]. When adjusting the value of $k_z$, the isofrequency contour for the bottom layer stays intact, but the isofrequency contour for the top layer, although maintaining its shape, is shifted by $A_y$=tan($\varphi$)$k_z$ in the $k_y$ direction. Hence, by selecting the value of $k_z$ we can obtain an effective gauge potential $\mathbf{A} = A_y \hat{y}$ in the reduced 2D system. More details of the metamaterial design can be found in the Supplemental Materials.

In the reduced 2D system, when an acoustic wave is incident from the bottom layer (henceforth called the "normal medium") to the top layer (henceforth called the "gauge field medium"), as shown in Fig. 1(b), the refraction is governed by a modified Snell's law involving the gauge field:

$$\sin(\theta_1)n_1 = \sin(\theta_a)n_2 + A_y / k_0 , \qquad (1)$$

$$\theta_2 = \arctan[\tan(\theta_a)n_x^2 / n_y^2] , \qquad (2)$$

with
$$n_2 = n_x n_y / \sqrt{[n_y \cos(\theta_a)]^2 + [n_x \sin(\theta_a)]^2} . \qquad (3)$$

Here $\theta_1$, $\theta_2$ are the angles of incidence and refraction, respectively, $n_2$ is an effective refractive index in the gauge field medium, and $\theta_a$ is the angle of the local wavevector in the gauge field medium [the green arrow in Fig. 1(b)]. Note that the gauge-field medium is slightly anisotropic, but the anisotropy alone does not cause negative refraction. Equation (1) is obtained from the continuity boundary condition, whereas Eq. (2) describes the relationship between the local wavevector and the group velocity in the anisotropic gauge field medium; Equation (3) indicates that the effective refractive index $n_2$ varies with $\theta_a$, owing to the anisotropy of the gauge-field medium. One can see that $\theta_a$ and $\theta_2$ have the same sign and $\theta_a = \theta_2$, when $n_x = n_y$ (i.e., the gauge-field medium becomes isotropic).

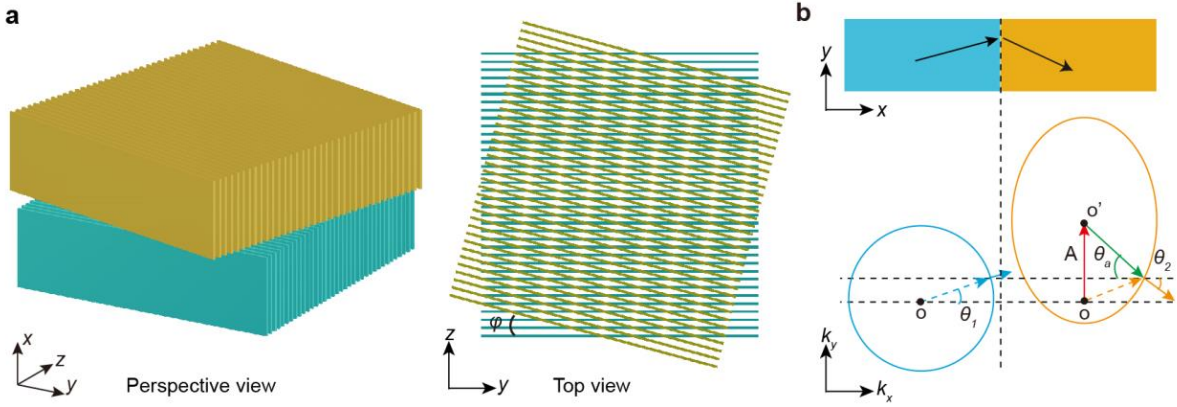

**Figure 1. Negative refraction induced by gauge fields in a twisted bilayer acoustic metamaterial.** (a) Perspective and top views of the acoustic metamaterial. The bottom (blue) and top (orange) layers are stacked hard solid plates in air (the background medium). The top layer is twisted around the *x*-axis by $\varphi$, forming a Moiré pattern interface with the bottom layer. (b) Synthetic gauge fields and negative refraction in the twisted bilayer acoustic metamaterial. For a fixed $k_z$, the gauge fields for bottom and top layers are 0 and $\mathbf{A}=\tan(\varphi)k_z\,\hat{y}$, respectively. The blue dashed (solid) arrow denotes the wavevector (group velocity) of the incident beam. The orange dashed (solid) arrow denotes the wavevector (group velocity) of the refracted beam. $\theta_1$, $\theta_a$, and $\theta_2$ are the incidence angle, the angle of the local wavevector in the gauge-field medium, and the refraction angle, respectively. The red and green arrows represent the gauge field and the local wavevector of the refracted beam, respectively.

We then proceed to demonstrate the phenomenon of gauge field induced negative refraction. The fabricated sample with $\varphi=45°$ is shown in Fig. 2(a). Acoustic waves are coupled into the sample via a rectangular waveguide with a narrow width in the *z*-direction, as shown in Fig. 2(b) and Fig. S2. The width along *z* is so narrow that the input can be treated as a delta function in *z*, whose Fourier transform covers almost all $k_z$ components. In the *x-y* plane, the waveguide launches an acoustic beam into the normal medium with incidence angle $\theta_i=15°$. We set the coordinate system such that the origin lies at the crossing point between the waveguide axis and the Moiré pattern interface. We then map out the output field pattern along the right surface (see Fig. S2). By locating the center of the output beam, we can determine the angle of refraction. By extracting different $k_z$ components, we obtain results for different gauge fields, as shown in Fig. 2(c). As the gauge field strength varies, the angle of refraction evolves from positive [$A_y = 0$ m$^{-1}$; see Fig. 2(e)], to zero [$A_y = 33$ m$^{-1}$; see Fig. 2(f)], and to negative [$A_y = 150$ m$^{-1}$; see Fig. 2(g)]. Note that the threshold value of $A_y$, corresponding to zero-angle refraction, matches the $k_y$ of the incident wave. After passing this threshold, the incident wave couples to the lower branch of the isofrequency contour of the gauge field medium, resulting in negative refraction. When the gauge field drags the isofrequency contour to other regions with no point matched to $k_y$ of the incident wave, total internal reflection occurs [$A_y = -200$ m$^{-1}$; see Fig. 2(h)].

These experimental observations can be explained in terms of the modified Snell's law. According to Eqs. (1)-(3), we have four different scenarios depending on the gauge-field potential, including positive refraction with $[\sin(\theta_1)k_0 - \sqrt{2}k_0] < A_y < \sin(\theta_1)k_0$ (-149 m$^{-1}$ < $A_y$ <33 m$^{-1}$ in experiment), zero-angle refraction with $A_y = \sin(\theta_1)k_0$ ($A_y$ = 33 m$^{-1}$ in experiment), negative refraction with $\sin(\theta_1)k_0 < A_y < [\sin(\theta_1)k_0 + \sqrt{2}k_0]$ (33 m$^{-1}$ < $A_y$ < 216 m$^{-1}$ in experiment), and total internal reflection with $A_y > [\sin(\theta_1)k_0 + \sqrt{2}k_0]$ or $A_y < [\sin(\theta_1)k_0 - \sqrt{2}k_0]$ ($A_y$ > 216 m$^{-1}$ or $A_y$ < -149 m$^{-1}$ in experiment). Based on Eqs. (1)-(3), we obtain the analytical solutions of the beam center positions at the measured plane, as shown by the green lines in Figs. 2(c)-(d). The measured, simulated and analytical results are in excellent agreement.

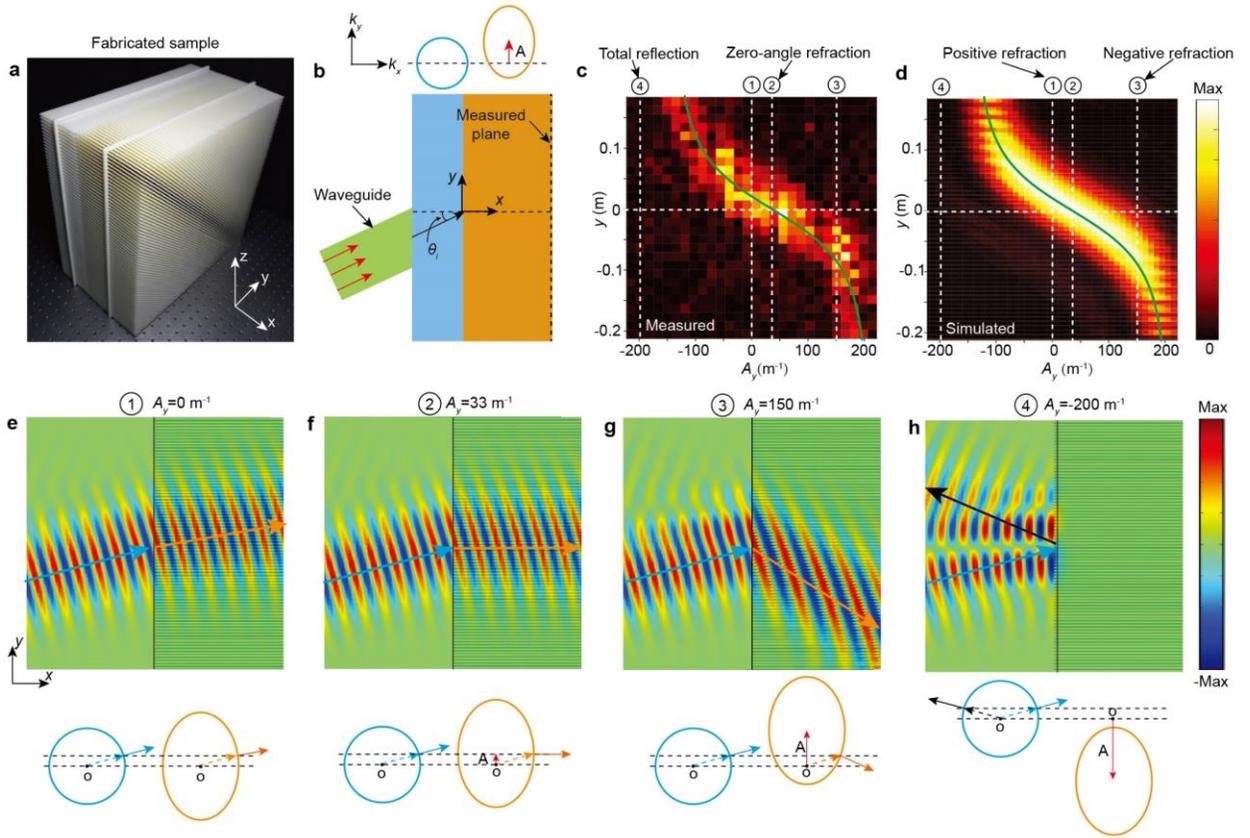

**Figure 2. Transition between positive and negative refraction with varying gauge field strength.** (a) Photograph of the fabricated sample. (b) Top view of the experimental setup. The width of the gauge-field medium is 140 mm. The acoustic field distributions are measured on the sample's right surface at 7 kHz. The acoustic wave is coupled into the sample via a rectangular waveguide. The incidence angle of the acoustic beam from the waveguide is $\theta_a$=15°. (c)-(d) Measured and simulated acoustic energy distributions at the right surface of the sample for different gauge field strengths. The green line represents the analytical solution for the beam center in the measured plane. The color bar indicates the energy intensity. (e)-(h) Simulated acoustic field distributions under different gauge field potentials

**A**. The plotted range for each figure is $[-200 \text{ mm}, 200 \text{ mm}]^2$. The color bar indicates the acoustic pressure. The bottom panels show momentum-space diagrams with dashed (solid) arrows denoting the wavevector (group velocity).

The gauge fields effectively break time-reversal (T) symmetry, rendering the in-plane beam propagation nonreciprocal[29]. It should be noted that the 3D acoustic metamaterial as a whole is T-symmetric, but in the reduced 2D system, for each specified nonzero $k_z$, T can be effectively broken. This is similar to how a T-symmetric Weyl semimetal can be projected into 2D by taking a fixed wavevector, whereupon it is described by a 2D Chern insulator with broken T[37,47,48].

To explicitly demonstrate this T-breaking, we set up the experiment shown in Fig. 3. An acoustic beam is incident from the gauge field medium to the normal medium, with incidence angle $\theta_i=6.6°$. The field pattern on the left surface at frequency 8 kHz is measured. Extracting the component with $A_y=137$ m$^{-1}$ [left column of Fig. 3(d)], we find the refracted beam center lies at around 0.1 m, indicating negative refraction. This measurement result is consistent with the simulated field pattern [right panel of Fig. 3(d)]. Next, we reverse the refraction beam and measure the field pattern along the right surface, as shown in Fig. 3(b). Extracting the component with $A_y=137$ m$^{-1}$, we observe total internal reflection, as shown in Fig. 3(e). Hence, the interface between the normal and gauge-field media acts as a one-way mirror. Both the nonreciprocal negative refraction and one-way mirror can be explained from momentum-space analysis, as shown in Fig. 3(c).

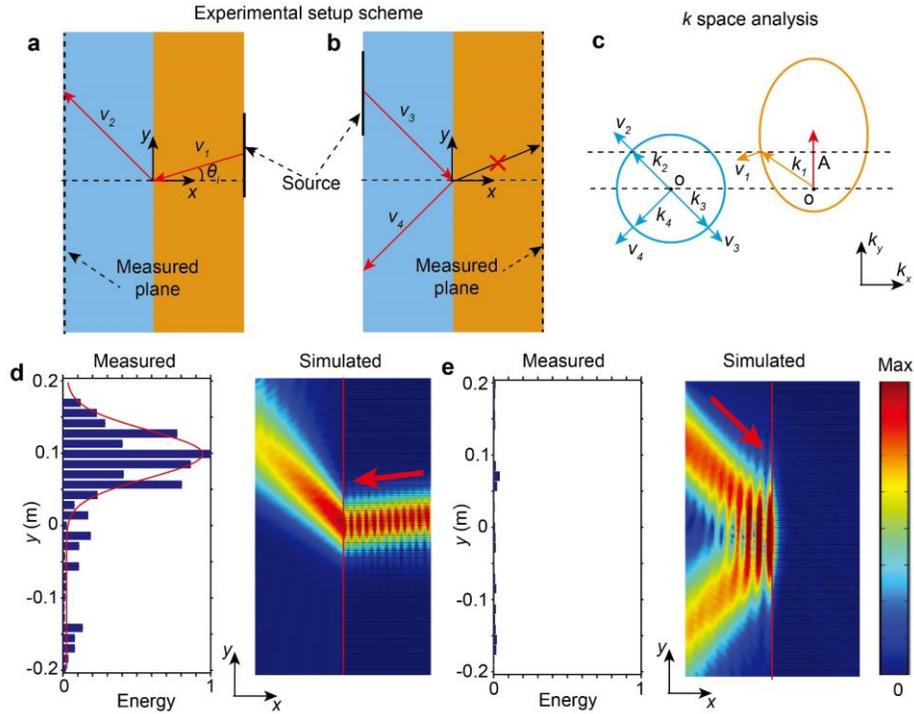

**Fig. 3. Observation of nonreciprocal negative refraction and one-way mirror.** (a)-(b) Top views of the experimental setup for different incident beams. The widths of the gauge field medium (orange) and the normal

medium (blue) are 120 mm. The field distributions at 8 kHz are measured along the sample's left (right) surface for incidence from the right (left) side. The incident angle $\theta_i$ is 6.6°. (c) Momentum-space analysis. $k_1$ ($k_2$) and $v_1$ ($v_2$) are the incident (refracted) momentum and group velocity of the beam. $k_3$ and $v_3$ are the time-reversed counterparts of the refracted beam associated with $k_2$. $k_4$ and $v_4$ are the reflected momentum and group velocity corresponding to the incident beam with $k_3$. (d)-(e) Left: normalized measured energy with $A_y$=137 m$^{-1}$ along the measurement plane indicated in (a) and (b). The red curve in the left inset of (d) is the fitted Gaussian profile of the measured energy distribution. Right: simulated field distributions, with colors representing the acoustic pressure amplitude.

The existence of total internal reflection can be further exploited to guide waves, as demonstrated recently in the context of a photonic lattice[26]. However, the drag effect caused by the gauge field can cause much richer physics than what has previously been explored. For example, similar to the transition from positive to negative refraction, we can cause the waveguide modes to undergo a transition between forward propagation (i.e., phase and energy propagating in the same direction) and backward propagation (i.e., phase and energy propagating in opposite directions)[20].

To accomplish this, we fabricated the twisted tri-layer structure shown in Fig. 4(a). At a fixed $k_z$, the cladding and core layers have gauge fields of 0 and $\mathbf{A}$= tan(45°)$k_z$ $\hat{y}$ , respectively, as shown in Fig. 4(b). A broadband loudspeaker is placed at the centre of the core layer to excite the guided modes. The field distributions in the *y-z* plane close to the Moiré pattern interface are mapped out (see Supplemental Material for details). Applying the Fourier transform to the measured field distributions, we obtain the dispersion curves in the $k_y$-$k_z$ momentum space over a range of frequencies. For a fixed gauge field strength (corresponding to a fixed $k_z$), the dispersion can be plotted as a function of $k_y$ and frequency. Thus, we experimentally characterize a class of acoustic gauge-field waveguides with different gauge field strengths, as shown in Figs. 4(c)-(i). For comparison, numerically simulated dispersions of the guided modes are plotted as white lines. The corresponding analytical calculations, based on the continuum theory, are shown in Fig. S3 of the Supplemental Materials.

As shown in Fig. 4(c), in the absence of gauge fields, the guided modes emerging from the light cone are symmetric around $k_y$ = 0 (these guided modes can exist because the light cone in the core is slightly larger than that in cladding layers). In presence of gauge fields, as shown in Figs. 4(d)-(i), we observe the following four striking features of gauge field induced waveguiding, consistent with earlier theoretical predictions[20]: (i) The dispersion curves of the guided modes are asymmetric around $k_y$=0, indicating nonreciprocity. (ii) The fundamental mode acquires a nonzero cut-off frequency. This is unlike conventional waveguides, where the fundamental dispersion curves generally start from zero frequency, like in Fig. 4(c). (iii) In certain frequency ranges, only one unidirectional mode exists, as for example in the blue-highlighted regions in Figs. 4(e)-(f). Single-mode one-way waveguiding is known to arise from the bulk band topology of Chern insulators (or their classical wave analogues)[30,32,38], but in this case the phenomenon arises in the context of a simple gauge field configuration without complicated band structure effects. (iv) Backward propagating modes with negative group velocities ($\partial f / \partial k_y < 0$), as shown in Figs. 4(g)-(i). From the momentum space diagrams of Fig. 4(b) and (j), it can be seen that these backward propagating modes emerge once the gauge field strength crosses a certain threshold value.

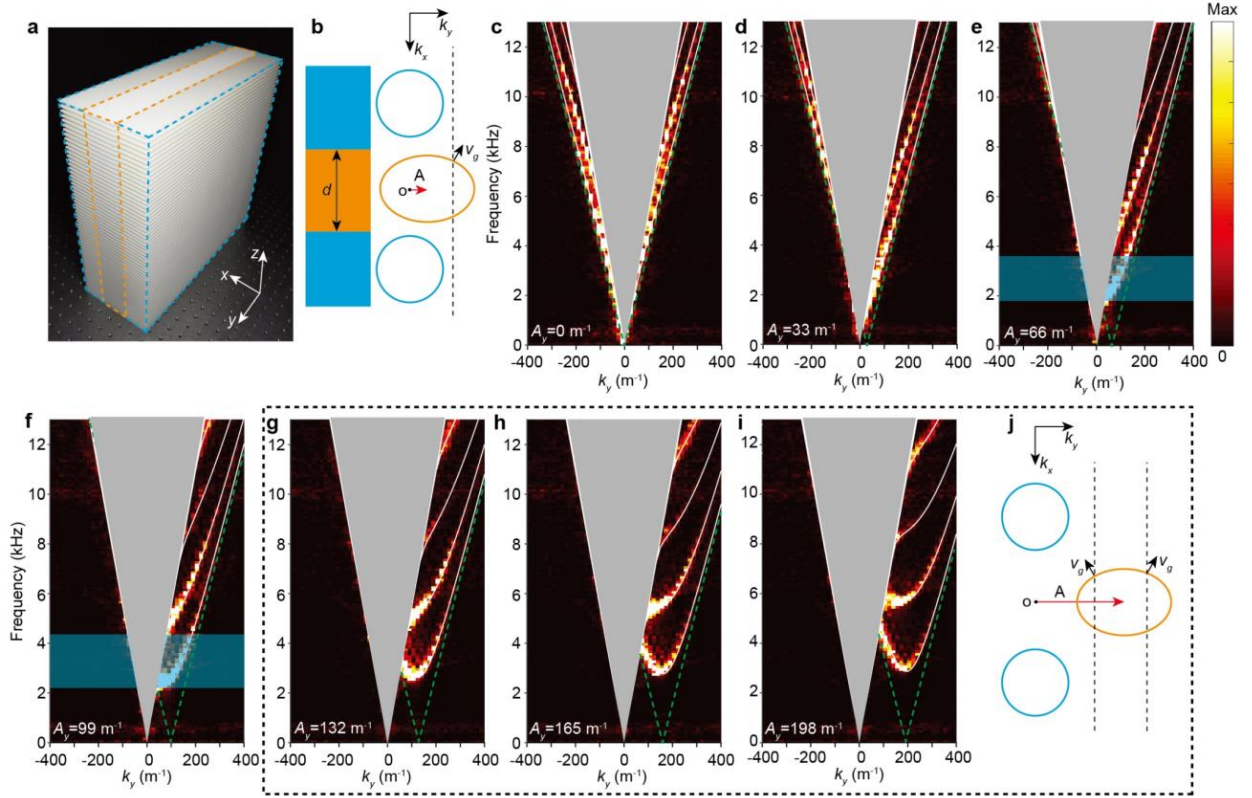

**Figure 4. Experimental demonstration of acoustic gauge-field waveguides with backward propagating modes.** (a) Photograph of the experimental sample, consisting of a twisted tri-layer structure. The blue and orange regions denote the layers without rotation and with rotation $\varphi=45°$, respectively. The thickness of the middle layer is $d$=50 mm. (b) Momentum-space analysis for small gauge fields, for which only guided modes with $v_g$>0 exist. (c)-(i) Measured dispersion of the gauge-field waveguides for different effective gauge field strengths. The color map represents the measured dispersion relation. The color bar at the right of (e) represents the energy intensity. The white lines are the numerically calculated dispersion curves. The green dashes and the grey regions represent the light cones for the core and cladding, respectively. The blue highlights indicate the region of single-mode one-way waveguiding. (j) Momentum-space analysis for large gauge fields, corresponding to the cases (g)-(i), for which there exist guided modes with $v_g$>0 and $v_g$<0.

We have thus experimentally demonstrated negative refraction induced by gauge fields. The designed twisted bilayer acoustic metamaterial is ideal for this demonstration due to having a straightforward continuum description, and continuous tunability of the synthetic gauge fields via the choice of out-of-plane wavenumber. Synthetic gauge fields are a promising way to produce a range of interesting wave phenomena, distinct from previously-studied approaches based on negative bulk refractive indices or phase discontinuities. We have demonstrated the use of gauge fields to create waveguides in which backward propagation characteristics emerge beyond a certain threshold gauge field strength. Considering the broad interest in metamaterials and metasurfaces in the past decades, we envision that this work may stimulate subsequent studies based on various physical settings including light and elastic waves.

## Acknowledgements

This work was sponsored by the Singapore Ministry of Education under Grants No. MOE2016-T3-1-006 and MOE2019-T2-2-085. H.S. acknowledges the support of the National Natural Science Foundation of China under grants No. 11774137 and 51779107, State Key Laboratory of Acoustics, Chinese Academy of Science under grant No. SKLA202016.